\documentclass[letter]{aa} 

\usepackage{graphicx}
\usepackage{txfonts}
\usepackage{hyperref}
\usepackage{mathrsfs}

\newcommand{\mrm}[1]{\mathrm{#1}}
\newcommand{\nuc}[2]{$\mrm{^{#2}#1}$}

\begin{document}

   \title{Vertical position of the Sun with $\gamma$-rays}

\author{
        Thomas Siegert
}
\institute{
        Center for Astrophysics and Space Sciences, University of California, San Diego, 9500 Gilman Dr, 92093-0424, La Jolla, USA
        \label{inst:ucsd}\\
        \email{tsiegert@ucsd.edu}
}
   \date{Received September 9, 2019; accepted October 16, 2019}
%
%
  \abstract
   {We illustrate a method for estimating the vertical position of the Sun above the Galactic plane by $\gamma$-ray observations. Photons of  $\gamma$-ray wavelengths are particularly well suited for geometrical and kinematic studies of the Milky Way because they are not subject to extinction by interstellar gas or dust. Here, we use the radioactive decay line of \nuc{Al}{26} at 1.809\,MeV to perform maximum likelihood fits to data from the spectrometer SPI on board the INTEGRAL satellite as a proof-of-concept study. Our simple analytic 3D emissivity models are line-of-sight integrated, and varied as a function of the Sun's vertical position, given a known distance to the Galactic centre. We find a vertical position of the Sun of $z_0 = 15 \pm 17$\,pc above the Galactic plane, consistent with previous studies, finding $z_0$ in a range between 5 and 29\,pc. Even though the sensitivity of current MeV instruments is several orders of magnitude below that of telescopes for other wavelengths, this result reveals once more the disregarded capability of soft $\gamma$-ray telescopes. We further investigate possible biases in estimating the vertical extent of $\gamma$-ray emission if the Sun's position is set incorrectly, and find that the larger the true extent, the less is it affected by the observer position. In the case of \nuc{Al}{26} with an exponential scale height of 150\,pc (700\,pc) in the inner (full) Galaxy, this may lead to misestimates of up to 25\,\%.}

   \maketitle
%

\section{Introduction}\label{sec:intro}

Measuring the vertical position of the Sun is interesting and important as it provides insights about Galactic kinematics, gravitational potentials, the history of the Solar System, and whether the Sun is located in a special position inside the Milky Way. \citet{Karim2016_Sun} described the need for an absolute reference frame of the Galactic Coordinate System, historically, and in the context of current$\mrm{ }$ microarcsecond-resolution astrometry. From their finding that the North Galactic Pole is $90.120^{\circ} \pm 0.029^{\circ}$ away from the dynamical centre of the Galaxy, that is, Sgr A*, the authors estimate that the Sun's position above the Galactic midplane is currently $17\pm5$\,pc to correct the absolute coordinate system. Thus, an accurate determination of the Sun's absolute position will provide a more stable Galactic Coordinate System, which in return will allow for more precise measurements of Galactic kinematics and its overall structure.

With the sub-percent measurement of the distance to the Galactic centre of $R_0 = 8178 \pm 35$\,pc by \citet{Abuter2019_GalCenDistance} {\citep[see also][finding $R_0 = 7946 \pm 82$\,pc]{Do2019_S0-2_galcen}}, it is now possible to move all the relative measurements, which typically use values between 7.5 and 9.0\,kpc, on common grounds, and also to use different methods to coherently determine the vertical height of the Sun above the disc, $z_0$. In most studies intending to determine $z_0$, the counts of specific astrophysical objects, such as stars \citep[e.g.][]{Elias2006_solarneighbourhood_zSun,Juric2008_SDSSstellarnumberdensity_zSun,Majaess2009_CepheidsMW_zSun}, globular or open clusters \citep[e.g.][]{Bonatto2006_openclusters_zSun,Buckner2014_starclusters_zSun,Joshi2016_openclusters_zSun} or magnetars \citep{Olausen2014_magnetars_zSun} and their spatial and kinematic distributions are used. Alternatively, H\,I \citep{Gum1960_HIGalaxy_zSun} or H\,II \citep[e.g.][]{Paladini2003_HIIcatalogue_zSun} regions, interstellar dust \citep[e.g.][]{Mendez1998_dustreddeningsolar_zSun}, or background star light \citep[e.g.][]{Freudenreich1998_BoxyBulge_COBE} provide similar estimates. A median of 56 measurements during the last century provides an estimate {for $z_0$ of $17$\,pc}, howver assuming different distances to the Galactic centre. {Recent estimates, including measurements of the past ten years, obtain $z_0$ {values between 5 and 29\,pc} \citep{Karim2016_Sun}, and show a large systematic spread among different as well as similar methods.}

{Measurements of the vertical position of the Sun above the Galactic plane mostly rely on the knowledge of an absolute reference frame: the IAU-accepted coordinate system from 1960 was defined with an accuracy of $\approx 0.1^{\circ}$ \citep{Blaauw1960_IAUcosy,Gum1960_HIGalaxy_zSun}. This value directly translates into an uncertainty for determining $z_0$ in absolute terms. The statistics and precision of current measurements allow for a reduction of this uncertainty to $\lesssim 0.03^{\circ}$, that is, to a level where systematic differences become important again \citep[e.g.][]{Peretto2009_darkSpitzerclouds,Simpson2012_galplanebubbles,Karim2016_Sun}. In addition, the assumption that the Galactic plane remains flat throughout may also introduce large systematic effects in various methods because it was shown that the Milky Way disc is warped as well \citep[e.g.][estimating $z_0 = 14.5 \pm 3.0$\,pc including a warp]{Skowron2019_MWwarp}.}

In this paper, we present a method for determining $z_0$ based on the observed emission morphology of the \nuc{Al}{26} $\gamma$-ray line at 1808.74\,keV. Depending on the Sun's vertical position, the full-sky appearance, and even more the emission peak, changes as a function of galactic latitude. Using analytic 3D emissivity models, in particular doubly exponential disc geometries, we derive line-of-sight integrated morphologies, and fit for $z_0$. In Sect.\,\ref{sec:method} we explain the general geometry as well as the fitting procedure. Our results are presented in Sect.\,\ref{sec:results}, followed by a study of possible biases in modelling the MeV sky latitudinally in Sect.\,\ref{sec:bias_study}. We conclude in Sect.\,\ref{sec:conclusion}.

\section{Modelling the $\gamma$-ray sky}\label{sec:method}

\subsection{\nuc{Al}{26} emission, line-of-sight integration, and geometry}\label{sec:los_geometry}

The radioactive nucleus \nuc{Al}{26} is ejected in winds of massive stars and their core-collapse supernovae \citep{Diehl2006_26Al}, and probably to a lesser extent in asymptotic giant branch stars and {classical} novae \citep{Diehl2018_ARI}, into the interstellar medium. With a {characteristic} lifetime of $\tau \approx 1.04$\,Myr, the nuclei experience $\beta^+$ decay to an excited state of \nuc{Mg}{26}, and almost instantaneously de-excite by the emission of a 1808.74\,keV $\gamma$-ray photon. At the time of the decay, the \nuc{Al}{26} ejecta have been distributed quasi-homogeneously around the massive stars in wind- and supernova-blown H\,I cavities \citep{Krause2014_FeedbackOrion}. As massive stars are aligned with the spiral arms of the Galaxy, it has been suggested from the kinematics of the \nuc{Al}{26} line \citep{Kretschmer2013_26Al}, that the global emission is found in a bubble-like structure preceeding the arms \citep{Krause2015_26Al}.

Direct imaging with the most modern $\gamma$-ray telescopes is not possible, and thus the physical parameters of a system cannot and should not be extracted by fits to an image. Instead, a (parametrised) model of the emission is to be convolved with the imaging response of the telescope, in order to perform maximum likelihood fits directly in the raw, photon-counting, and instrument-specific data space.

\begin{figure}[!th]
        \centering
        \includegraphics[width=\columnwidth,trim=0.5in 0.5in 0.5in 0.5in,clip=true]{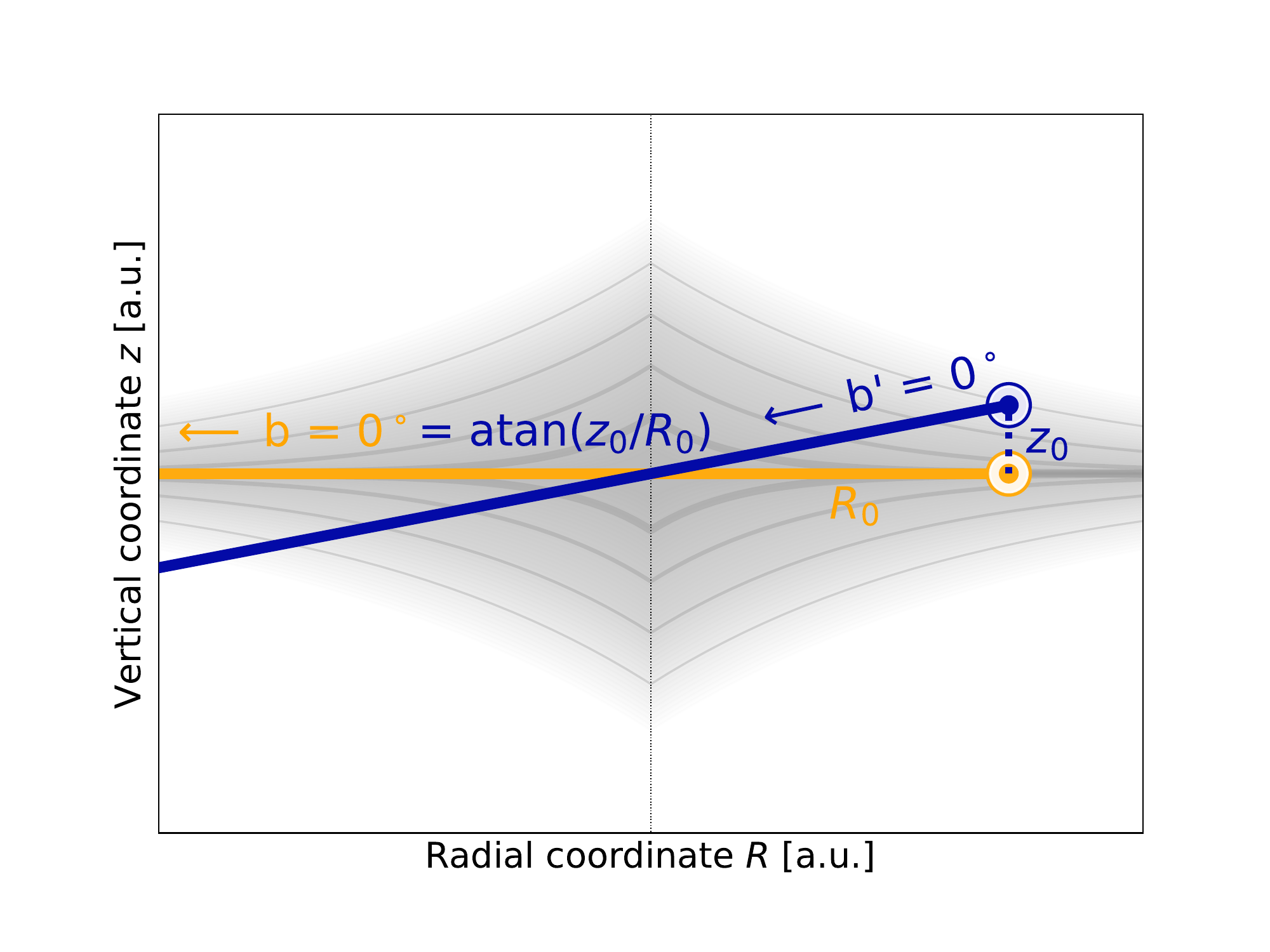}
        \caption{Sketch illustrating the different line-of-sight integration paths for the same apparent latitude angle, $b$/$b'$, as determined by the vertical position of the Sun, $z_0$, at fixed radial position, $R_0$.}
        \label{fig:sketch}
\end{figure}

While iterative methods, such as the Richardson-Lucy deconvolution \citep[e.g.][]{Knoedlseder2005_511} or the maximum entropy method \citep[e.g.][]{Bouchet2015_26Al}, have been used to produce images in the $\gamma$-ray domain, their interpretation is on shaky grounds because the algorithms can suffer from a lack of objectivity \citep[see, e.g.,][]{Allain2006_gammaimaging}. To determine physical parameters, a full forward-modelling approach, including the dominating instrumental background, the imaging response, and a geometrical or kinematic model should be preferred. With more data, details in the emission morphology are of course revealed. It nevertheless has been shown in many previous studies \citep[e.g.][]{Knoedlseder1996_26Al,Oberlack1996_26Al,Diehl2006_26Al,Wang2009_26Al,Kretschmer2011_PhD,Kretschmer2013_26Al,Siegert2017_PhD} that to large extents, a simple three-parameter model can explain the full-sky \nuc{Al}{26} emission at 1.8\,MeV, a doubly exponential disc:

\begin{equation}
\rho(x,y,z) = \rho_0 \exp\left(-\frac{R}{R_e}-\frac{|z|}{z_e}\right)\mrm{.}
\label{eq:exp_disk}
\end{equation}

In Eq.\,(\ref{eq:exp_disk}), $\rho(x,y,z)$ is the instantaneous photon emissivity in 3D space, with ${(x,y,z) = (0,0,0)}$ resembling the Galactic centre, in units of $\mrm{ph\,cm^{-3}\,s^{-1}}$, $\rho_0$ is the normalisation, inherent to the total galactic 1.8\,MeV luminosity and the received or measured flux at the point of the observer, $R_e$ and $z_e$ are the exponential scale radius and scale height, respectively, and $R = \sqrt{x^2 + y^2}$ is the Galactocentric radius.

In order to obtain an image from the the distribution in Eq.\,(\ref{eq:exp_disk}), we performed line-of-sight integrations, Eq.\,(\ref{eq:los_integration}), starting from the position of the observer, here the Sun with ${(x_0,y_0,z_0)}$, outwards. This requires a coordinate transformation,

\begin{eqnarray}
& x'(s) & = x_0 - s \cos l \cos b \nonumber\\
& y'(s) & = y_0 - s \sin l \cos b\nonumber\\
& z'(s) & = z_0 - s \sin b\mrm{,}
\label{eq:los_trafo}
\end{eqnarray}

with $s$ being the sight-line in galactic-coordinate direction $({l,b})$, such that

\begin{eqnarray}
& R'^2(s) & = x'^2(s) + y'^2(s) = R_0^2 + s^2 \cos^2 b - 2sp_0\mrm{,}
\label{eq:R_trafo}
\end{eqnarray}

and $p_0 = \cos b \cos l + \cos b \sin l$. The line-of-sight integration thus reads

\begin{equation}
F_{los}({l,b}) = \frac{\rho_0}{4\pi} \int_{s_{min}}^{s_{max}} ds \exp\left(-\frac{R'(s)}{R_e}\right)\exp\left(-\frac{|z'(s)|}{z_e}\right)\mrm{,}
\label{eq:los_integration}
\end{equation}

\begin{figure}[!ht]
        \centering
        \includegraphics[width=\columnwidth,trim=0.1in 0.1in 0.5in 0.5in,clip=true]{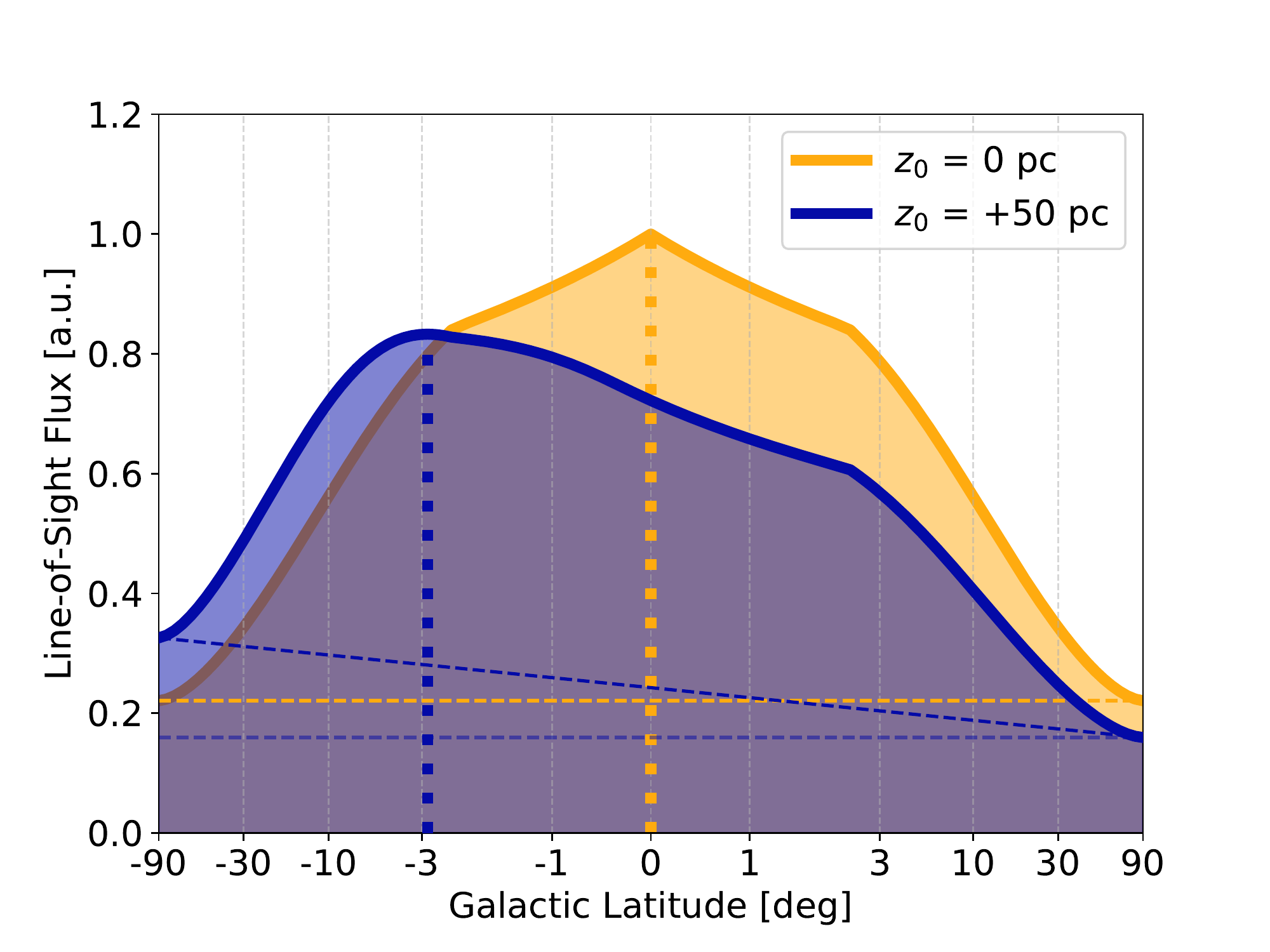}
        \caption{Longitude-averaged line-of-sight integration of the same exponential disc ${(R_e,h_e) = (3.5,0.1)}$ 3D emissivity model as a function of latitude for vertical positions of 0 (symmetric, orange) and $+50$\,pc (violet). In the symmetric case, the emission peaks at $b=0^{\circ}$, as expected, and already for $z_0 = 50$\,pc, the peak is shifted $3^{\circ}$ to negative latitudes. See text for further details and implications.}
        \label{fig:los-integration}
\end{figure}
\begin{figure*}[!th]
        \centering
        \includegraphics[width=\textwidth,trim=0.8in 1.0in 0.8in 0.8in,clip=true]{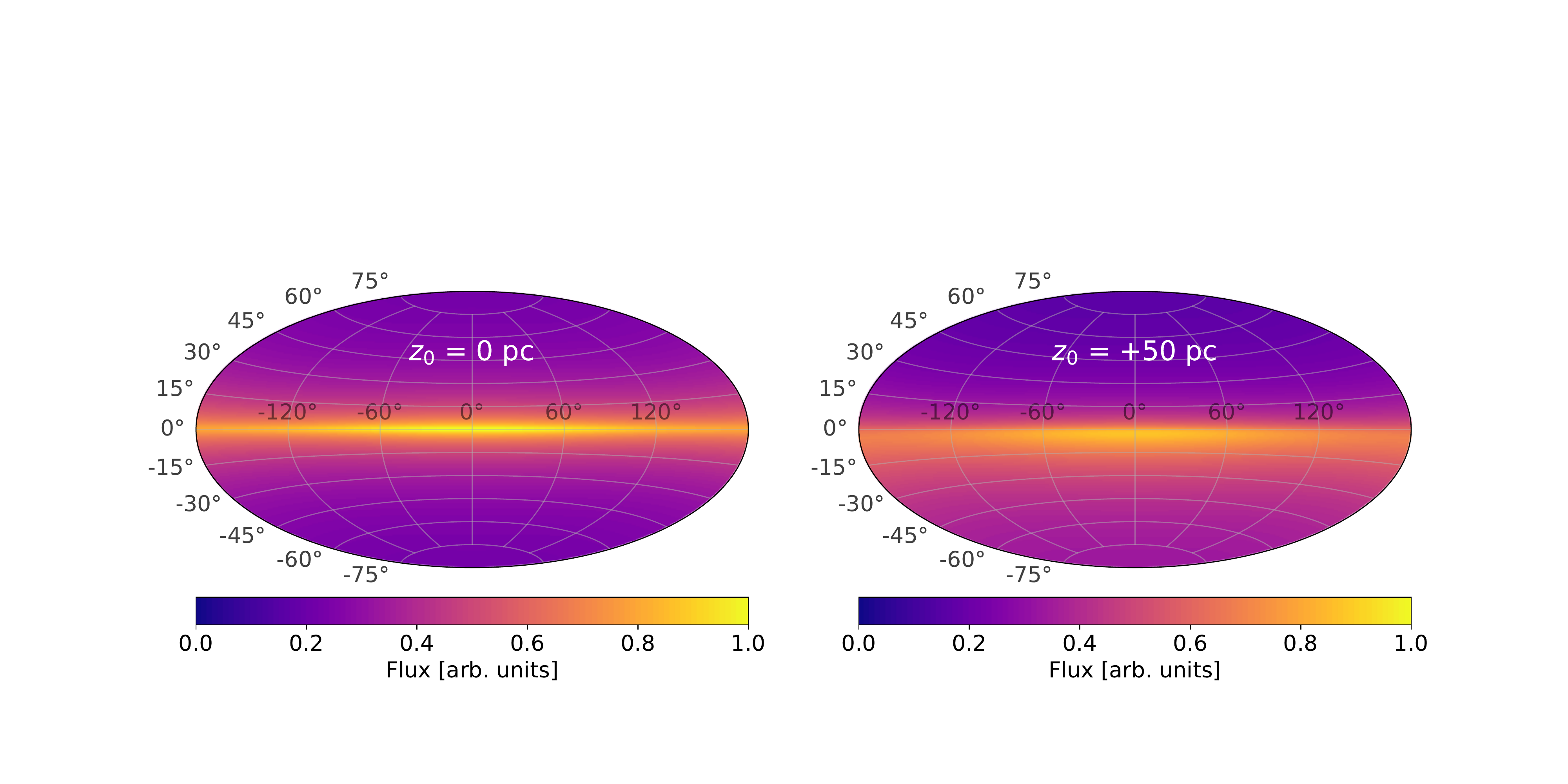}
        \caption{Line-of-sight integrated emissivities of the same exponential disc ${(R_e,h_e) = (3.5,0.1)}$ model as a function of longitude and latitude for $z_0=0$ (left) and $+50$\,pc (right). The images have been normalised to the symmetric case and thus contain the same absolute flux. Clearly, the contrast in the right image is higher, but the flux is more spread out into a larger number of pixels.}
        \label{fig:full-skies}
\end{figure*}


where $F_{los}({l,b})$ is the flux at longitude {and} latitude direction, $({l,b})$, in units of $\mrm{ph\,cm^{-2}\,s^{-1}\,sr^{-1}}$, and which is not analytically solvable. For a fixed distance to the Galactic centre, $R_0$, and each combination of the scale dimension, ${(R_e,z_e)}$, $F_{los}({l,b})$ becomes a function of the observer position, $z_0$, only. Geometrically, changing $z_0$ is equivalent to changing the coordinate frame latitudinally (see Fig.\,\ref{fig:sketch}), {resulting in an apparent `tilt' of the observed direction}. As a result, the former ($z_0=0$) symmetric emission profile, averaged over all longitudes, is skewed towards the opposite direction. In Fig.\,\ref{fig:los-integration}, the latitudinal emission profiles of the same exponential disc model with ${(R_e,z_e) = (3.5,0.1)}$ are shown for the symmetric case, and an observer position of $z_0 = +50$\,pc. Clearly, the emission peak is shifted towards negative latitudes if the observer is above the disc, here to about $b=-3^{\circ}$. It is important to note that the flux in both examples is the same, but more spread out to higher latitudes in the case of an observer outside the disc. In addition, there is a certain level of `isotropic emission' in each of the cases (horizontal lines), which especially for coded-mask telescopes in the soft $\gamma$-ray regime is nearly impossible to detect\footnote{We note that Compton telescopes are {barely} affected by this circumstance because they locate the emission based on the more unique Compton scattering response. See e.g. the Compton Spectrometer and Imager, COSI \citep{Kierans2016_COSI,Tomsick2019_COSI}.}. A slightly skewed morphology already reduces this isotropic part and generally allows for more precise measurements in the MeV range. The full-sky appearance of the two examples is shown in Fig.\,\ref{fig:full-skies}.

These detailed peculiarities can be used to determine $z_0$ from $\gamma$-ray emission alone. In the following Sect.\,\ref{sec:fit}, the basics about the spectrometer on board the International Gamma-Ray Astrophysics Laboratory (INTEGRAL/SPI), the chosen data set, and model fitting in its raw photon-count data space are explained.

\subsection{Instrument, data, and fit}\label{sec:fit}

The spectrometer SPI \citep{Vedrenne2003_SPI} on board ESA's INTEGRAL satellite \citep{Winkler2003_INTEGRAL} measures photons in the energy range between 20 and 8000\,keV with a 19-element high-purity Ge detector camera. It uses a coded-mask technique to distinguish between emission from the sky and typically high instrumental background radiation from the telescope and satellite themselves. SPI has a field of view of $16^{\circ}$ and an angular resolution of $2.7^{\circ}$. The data space is thus a size-19 vector of photon counts at different energies for individual pointed observations (`pointings') of typically 30\,min duration. Because we study one of the strongest $\gamma$-ray lines, we summed all counts between 1805 and 1813\,keV into one 8\,keV broad energy bin, corresponding to three times the spectral resolution of SPI at 1.8\,MeV. In this study, we do not aim for optimising or characterising the \nuc{Al}{26} emission morphology, nor are we interested in the detailed spectral shape of the emission line. 

Our data set consists of 200\,Ms of INTEGRAL/SPI observation time, distributed predominantly in the centre of the Galaxy as well as the Galactic plane, with a patchy exposure at higher latitudes\footnote{See \url{http://www2011.mpe.mpg.de/gamma/instruments/integral/spi/www/public_data/index.html} for a current exposure map.}. In total, more than 13 years of data are accumulated, resulting in $N_{obs} = 92,867$ pointings of consecutive size-19 vectors. When the imaging response, $R_{jp}$, is applied for each pixel $j \in ({l,b})_j$ in each pointing $p \in N_{obs}$, to the emission model $F_{los}({l,b})_j$, this translates an image or model of the sky, $M^{SKY}_p$, into the SPI data space for a particular set of observations:

\begin{equation}
M^{SKY}_p = F_{norm} \sum_j R_{jp} F_{los}({l,b};z_0)_j\mrm{.}
\label{eq:sky_model}
\end{equation}

In Eq.\,(\ref{eq:sky_model}), the absolute flux normalisation, $F_{norm}$, is determined through a maximum likelihood fit, and $z_0$ is a free parameter (see below). Understanding the instrumental background in $\gamma$-ray astrophysics is key to any parameter inference. Details about robust and reliable background modelling with SPI can be found in \citet{Diehl2018_BGRDB} and \citet{Siegert2019_SPIBG}. Here, the background is modelled as a combination of contributions from the instrumental continuum, $M^{BG,c}_p$, as well as instrumental lines, $M^{BG,l}_p$, and their absolute levels are determined through a (time-dependent) scaling parameter in a fit again. The total model thus writes

\begin{equation}
M^{tot}_p = F_{norm} \sum_j R_{jp} F_{los}({l,b};z_0)_j + \beta^c_t M^{BG,c}_p + \beta^l_t M^{BG,l}_p \mrm{.}
\label{eq:modeldesc}
\end{equation}

In this full forward-modelling manner, the parameters $F_{norm}$, $\beta^c_t$, and $\beta^l_t$ are straightforwardly determined by maximising the Poissonian likelihood function,

\begin{equation}
\mathscr{L}(D|M) = \prod_p \frac{M_p^{D_p} \exp(-M_p)}{D_p!}\mrm{,}
\label{eq:poisson_likelihood}
\end{equation}

because they are linear parameters. However, the convolution of $F_{los}({l,b};z_0)_j$ with $R_{jp}$ is computationally very expensive and would be required in each iteration of the fit to determine the unknown parameter $z_0$ because it is changing the appearance of $M^{SKY}_p$. We therefore fixed it for a particular fit, and evaluated the likelihood for a pre-defined set of $z_0$ values. Here, we used a grid of 15 vertical extents, ranging between $-20$ and $+130$\,pc in steps of 10\,pc. These values are based on a canonical distance of 8.5\,kpc to the Galactic centre. Because the problem is geometrically similar, a correction factor $R_0/\mrm{8.5\,kpc} = 0.9621$ can be applied to each $z_0$ value for a correct estimate given the more precise measurements of $R_0$ by \citet{Abuter2019_GalCenDistance}, {or $0.9348$ from the $R_0$ estimate by \citet{Do2019_S0-2_galcen}}.

\section{Results}\label{sec:results}

We used two examples to estimate the vertical position of the Sun from emission of the \nuc{Al}{26} decay line. The emission in both cases is characterised by the doubly exponential disc model, Eq.\,(\ref{eq:exp_disk}), but with two assumptions on the radial and vertical scale parameters: \citet{Wang2009_26Al} suggested an exponential scale height at 1.8\,MeV of $130_{-70}^{+120}$\,pc at a distance to the Galactic centre of 8.5\,kpc. This estimate is based on a fixed scale radius of 3.5\,kpc and only considers the sky region $|l|<60^{\circ}$, $|b|<30^{\circ}$. \citet{Siegert2017_PhD} used the data set described in this work and performed exponential disc fits, including the full sky and varying both scale dimensions. This resulted in an estimate of $R_e = 5.6 \pm 0.6$\,kpc and $z_e = 670 \pm 190$\,pc for \nuc{Al}{26}. We scaled and re-weighted the parameter sets of \citet{Wang2009_26Al} and \citet{Siegert2017_PhD}, resulting in the combinations ${(R_e,z_e) = (3.37,0.15)}$ and ${(4.81,0.46)}$, respectively. In {a} common reference system with $R_0 = 8.178$\,kpc, we finally performed the above-described maximum likelihood fits. These two examples serve as a proof-of-concept study that uses the two extreme values reported in the recent literature.

\begin{figure}[!ht]
        \centering
        \includegraphics[width=\columnwidth,trim=0.2in 0.2in 0.6in 0.2in,clip=true]{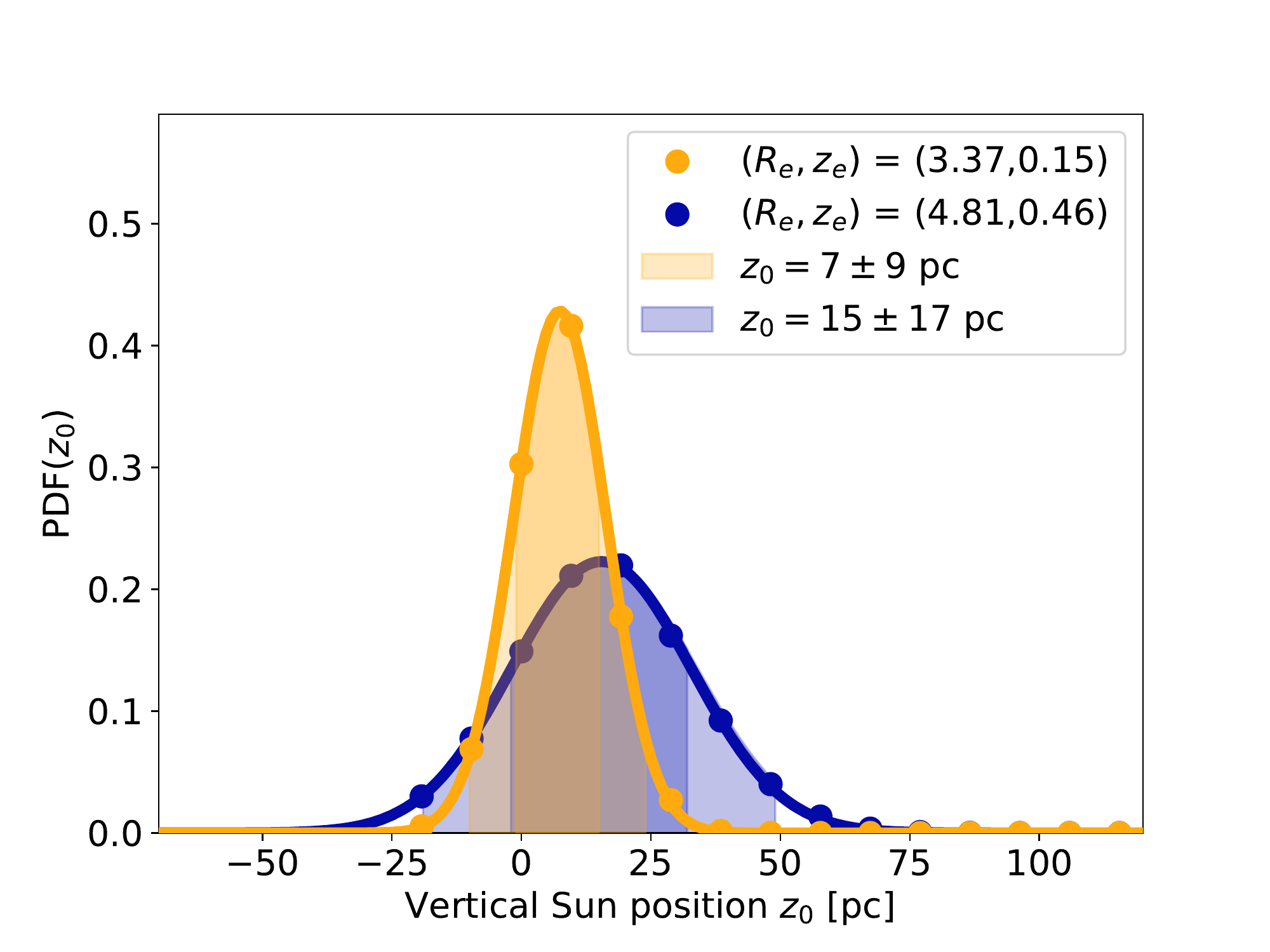}
        \caption{Probability distribution functions for the vertical position of the Sun using INTEGRAL/SPI \nuc{Al}{26} data and two different exponential disc models. In this range of plausible models, the position is constrained between in the interval [-2,32]\,pc, consistent with independent estimates.}
        \label{fig:SPI_z0}
\end{figure}

The probability density functions of $z_0$, given the two models, are shown in Fig.\,\ref{fig:SPI_z0}. The more concentrated emission model with small-scale dimensions results in a vertical position of the Sun above the Galactic midplane of $7 \pm 9$\,pc. The larger-scale dimensions, as suggested from the full-sky model fits, result in $z_0 = 15 \pm 17$\,pc. These values are consistent with each other and also consistent with the median of 56 previous completely independent measurements using different techniques, {giving $\approx 17$\,pc in a range between 5 and 29\,pc} \citep{Karim2016_Sun}. While the model with a narrower emission morphology provides smaller uncertainties, it was shown that larger scale radii and heights are apparently needed to cover the entire\footnote{Combinations of small and large scale heights provide an estimate of how much emission comes from the solar vicinity \citep{Siegert2017_PhD}, but this is very degenerate in the SPI data space and was thus not considered here as a third case.} Milky Way emission at 1.8\,MeV \citep{Siegert2017_PhD}. We therefore quote the larger uncertainty as a more conservative estimate. A more detailed model to describe the true \nuc{Al}{26} morphology in the Galaxy might even now result in a much more accurate estimate of $z_0$. The structural details of the \nuc{Al}{26} emission in the Milky Way are beyond the scope of this paper because the goal is to show the feasibility of such a measurement with currently available data in the soft $\gamma$-ray regime

\section{Vertical emission extents and bias}\label{sec:bias_study}

This remarkable and unexpectedly good result leads to the question whether there might be
biases in estimating the spatial extents of MeV $\gamma$-ray emission. In particular, setting the vertical position of the observer incorrectly in a typical analysis, for example, intending to measure the exponential scale height, may result in over- or underestimating the vertical extent of the emission. 

\begin{figure}[!ht]
        \centering
        \includegraphics[width=\columnwidth,trim=0.2in 0.2in 0.6in 0.2in,clip=true]{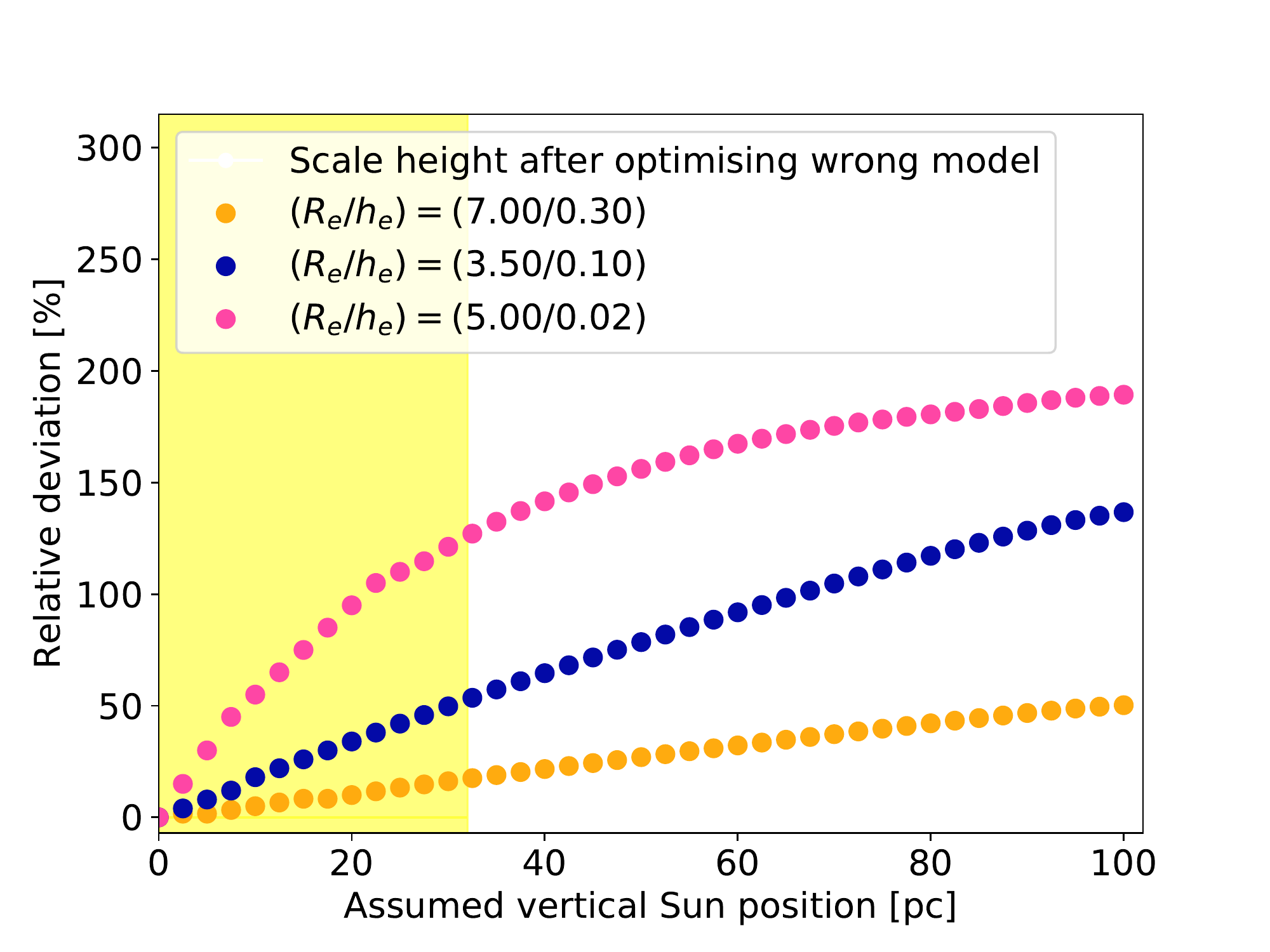}
        \caption{Relative deviations, $\mrm{True-Fitted}/\mrm{True} \times 100\,\%$, of doubly exponential and change throughout for consistency disc models as a function of $z_0$ and $z_e$ ($z_e = 300$\,pc orange, 100\,pc violet, and 20\,pc pink), fitted by a $z_0 = 0$ model, assuming the correct $R_e$. The yellow region determines the $1\sigma$ uncertainty from the $z_0$ fit from 1.8\,MeV data alone, see Sect.\,\ref{sec:results}. By definition, the $z_0 = 0$ case yields a relative deviation {of} zero.}
        \label{fig:scale_height_bias}
\end{figure}

We investigated this potential bias by fitting exponential disc models with $z_0 = 0$ to simulations of models with $z_0 \neq 0${, and} varying the scale height $h_e$. In this way, we determined an optimal scale height that differs from the input model if the correct position of $z_0$ is not met. We chose three different sets of ${(R_e,z_e)}$, ${(7.00,0.30)}$, ${(3.50,0.10)}$, and ${(5.00,0.02)}$, and performed a scan of $h_e$ for each value of $z_0$. The resulting relative deviations for our three test cases are shown in Fig.\,\ref{fig:scale_height_bias}. If the vertical position $z_0 = 0$ is met, the relative deviation from the correct scale height is by definition equal to zero. Because this problem is again symmetric, we only show positive values of $z_0$. The larger the true scale height, the smaller the relative deviation from using an incorrect observer position. In the case of ${(R_e,z_e) = (7.00,0.30)}$, for example, the relative deviation inside the $z_0$-position interval determined from $\gamma$-rays is up to 20\,\%. This value is comparable to the uncertainty of the \nuc{Al}{26} scale height determined by \citet{Siegert2017_PhD} ($<30\,\%$) and should therefore be carefully considered when new or updated measurements in the MeV regime are discussed. In the case of smaller true scale heights, the effect of choosing an incorrect observer position is even more severe and ranges up to 50\,\% (125\,\%) when scale heights as small as 100\,pc (20\,pc) are considered.

\section{Conclusion}\label{sec:conclusion}

Using 13 years of \nuc{Al}{26} data at 1.8\,MeV from INTEGRAL/SPI, we determined the vertical position of the Sun above the Galactic midplane to $z_0 = 15 \pm 17$\,pc in a proof-of-concept study. This is a remarkable result, both in terms of accuracy and precision: current soft $\gamma$-ray telescopes suffer from the `MeV gap' in sensivity \citep[e.g.][]{DeAngelis2018_eAstrogam,Timmes2019_RA2020,Tomsick2019_COSI}, which most of the time allows them to only study the {sources} inside, or the Milky Way itself. The {statistical} uncertainties on $z_0$ we estimate from only using INTEGRAL/SPI data and a simple first-order analytic geometric model are only a factor of a few away from the most precise measurements using star counts or H\,II regions ($17 \pm 2$\,pc){, which show a systematic spread of  $\approx 12$\,pc \citep{Karim2016_Sun}, however.} A more detailed geometric model of \nuc{Al}{26} would already provide similar uncertainties in $z_0$ with current instrumentation. In consequence, the accuracy of this measurement can imply biases in determining other emission parameters, such as its vertical extent (latitudinal or exponential scale height). We find that specifically for \nuc{Al}{26}, the relative uncertainty of scale heights as a function of $z_0$ can be up to 25\,\% (or higher for smaller scale heights than \nuc{Al}{26}). This value is comparable to the uncertainties from scale-height measurements themselves, and may lead to misestimates (in both directions) of the true scale height. Consequently, $z_0$ should be carefully considered in future modelling of the MeV $\gamma$-ray sky.

\begin{acknowledgements}
        Thomas Siegert is supported by the German Research Society (DFG-Forschungsstipedium SI 2502/1-1).
\end{acknowledgements}

\bibliographystyle{aa} 
\bibliography{alles} 

\end{document}